\documentclass[10pt]{article}

\usepackage{overcite}

\begin{document}

\markboth{T. Sauer}{The Einstein-Vari\'cak correspondence on relativistic rigid rotation}


\title{The Einstein-Vari\'cak Correspondence on Relativistic Rigid Rotation%
	\footnote{To appear in: Proceedings of the Eleventh Marcel 
	Grossmann Meeting on General Relativity, ed.~H.~Kleinert, R.T.~Jantzen
	and R.~Ruffini, World Scientific, Singapore, 2007.}}

\author{Tilman Sauer\\
        {\small Einstein Papers Project}\\[-0.1cm]
        {\small California Institute of Technology 20-7}\\[-0.1cm]
        {\small Pasadena, CA 91125, USA}\\[-0.1cm]
        {\small tilman@einstein.caltech.edu}}
\date{}

\maketitle

\begin{abstract}
The historical significance of the problem of relativistic rigid rotation is reviewed
in light of recently published correspondence between Einstein and the mathematician
Vladimir Vari\'cak from the years 1909 to 1913.
\end{abstract}

\section{Introduction}\label{intro}

The rigidly rotating disk has long been recognized as a
crucial `missing link' in our historical reconstruction of Einstein's
recognition of the non-Euclidean nature of spacetime in his
path toward general relativity.\cite{Stachel1980,MO1995}
Relativistic rigid rotation combines several different but related problems: the issue
of a Lorentz-covariant definition of rigid motion, the number of degrees
of freedom of a rigid body, the reality of length contraction,\cite{CPAE3} as well as
Ehrenfest's paradox\cite{Klein1970} and the introduction of non-Euclidean
geometric concepts into the theory of relativity.\cite{ZN}

\section{Relativistic rigid motion}

A relativistic definition of rigid motion was first given by Max Born.\cite{Born1909}
The definition was given in the context of a theory of
the dynamics of a model of an extended, rigid electron, and 
defined a rigid body as one whose infinitesimal volume elements appear undeformed for
any observer that is comoving instantaneously with the (center of the) respective volume element.
The definition and its implications were discussed at the 81st meeting of the Gesellschaft 
Deutscher Naturforscher und \"Arzte in Salzburg in late September 1909.

Gustav Herglotz and Fritz Noether,  in papers
received by the {\it Annalen der Physik} on 7 and 27 December, respectively, further
elaborated on the mathematical consequences of Born's definition.\cite{HerglotzNoether} 
Herglotz, in particular, 
reformulated the definition in more geometric terms: A continuum performs rigid motion if the 
world lines of all its points are equidistant curves. The analysis showed that Born's infinitesimal 
condition of rigidity can only be extended to the motion of a finite continuum in special cases.
It implied that a rigid body has only three degrees of freedom. The motion of one
of its points fully determines its motion. Translation and uniform rotation are special cases.
In particular, the definition does not allow for {\it acceleration} of a rigid disk from rest to
a state of uniform rotation with finite angular velocity.

In view of these consequences, various other definitions of a rigid body were
suggested, e.g.\ by  Born and Noether,\cite{HerglotzNoether,Born1910} until it
became clear that special relativity does not allow for the usual concept of a rigid body.
In other words, a relativistic rigid body necessarily has an infinite
number of degrees of freedom.\cite{EinsteinLaue}

On 22 November 1909, a short note appeared by Paul Ehrenfest pointing to a paradox that follows
from Born's relativistic definition of rigid motion of a continuum.\cite{Ehrenfest1909} 
He considered a rigid cylinder rotating 
around its axis and contended that its radius would have to meet two contradictory requirements.
The periphery must be Lorentz-contracted, while its diameter
would show no Lorentz contraction. The difficulty became known as the ``Ehrenfest paradox.''
In a polemic exchange with von Ignatowsky,\cite{Ignatowsky} Ehrenfest  
devised the following thought experiment to illustrate the difficulty. He imagined the  
rotating disk to be equipped with markers along the diameter and the periphery. If their positions
were marked onto tracing paper in the rest frame at a fixed instant, with the disk both at rest and 
in uniform rotation, the two images should show the same radius but different circumferences.

\section{The Einstein-Vari\'cak correspondence}

Immediately after the 1909 Salzburg meeting, Einstein wrote to Arnold Sommerfeld that ``the treatment
of the uniformly rotating rigid body seems to me of great importance because of an extension of
the relativity principle to uniformly rotating systems.''\cite{CPAE5} This was a necessary
step for Einstein following the heuristics of his equivalence hypothesis, but only
in spring 1912, a few weeks before he made the crucial transition from a scalar to a tensorial
theory of gravitation based on a general spacetime metric,\cite{ZN} do we find another
hint at the problem in his writings.\cite{Stachel1980,MO1995}

The {\it Collected Papers of Albert Einstein} recently published\cite{CPAE10} nine
letters by Einstein
to Vladimir Vari\'cak (1865--1942), professor of mathematics at Agram (now Zagreb, Croatia).
Vari\'cak had published on non-Euclidean geometry\cite{Varicak}
and is known for representing special relativistic relations in terms of real hyperbolic
geometry.\cite{Varicak1910,Walter1999} 
The correspondence seems to have been initiated by Vari\'cak asking for offprints of
Einstein's papers. In his response, Einstein added a personal tone to it with his wife 
Mileva Mari\'c, a native Hungarian Serb, writing the address in Cyrillic script in order
to raise Vari\'cak's curiosity. After exchanging publications,
Vari\'cak soon commented on Einstein's (now) famous 1905 special relativity paper, pointing to 
misprints but also raising doubts about his treatment of reflection of light rays off moving
mirrors. These were rebutted by Einstein in a response of 28 February 1910 in which he also, 
with reference to Ehrenfest's paradox, referred to the rigidly rotating disk
as the ``most interesting problem'' that the theory of relativity would presently have to offer.
In his next two letters, dated 5 and 11 April 1910 respectively, Einstein
argued against the existence of rigid bodies invoking the impossibility of
superluminal signalling, and also discussed the rigidly rotating disk. 
A resolution of Ehrenfest's paradox, suggested by Vari\'cak,
in terms of a distortion of the radial lines so as to preserve the ratio of $\pi$
with the Lorentz contracted circumference, was called interesting but not viable.
The radial and tangential lines 
would not be orthogonal in spite of the fact that an inertial observer comoving
with a circumferential point would only see a pure rotation of the disk's neighborhood. 

About a year later, Einstein and Vari\'cak corresponded once more. Vari\'cak had 
contributed to the polemic between Ehrenfest and von Ignatowsky by suggesting a
distinction between `real' and `apparent'
length contraction. The reality of relativistic length contraction was
discussed in terms of Ehrenfest's tracing paper experiment, but for linear relative motion.
According to Vari\'cak, the experiment would show that the contraction is only a psychological
effect whereas Einstein argued that the effect will be observable in the distance of the
recorded marker positions. When Vari\'cak published his note, Einstein responded with a 
brief rebuttal.\cite{VaricakEinstein}

Despite their differences in opinion, the relationship remained friendly.
In 1913, Einstein and his wife thanked Vari\'cak for sending them a gift, commented
favorably on his son who stayed in Zurich at the time, and Einstein announced
sending a copy of his recent work on a relativistic theory of gravitation.
The Einstein-Vari\'cak correspondence thus gives us additional insights into a 
significant debate. It shows Einstein's awareness of the intricacies of relativistic
rigid rotation and bears testimony to the broader context of the conceptual clarifications
in the establishment of the special and the genesis of the general theory of relativity.

\vfill
\end{document}